\documentclass[preprint,12pt]{elsarticle}
\usepackage{amssymb}
\usepackage{tikz}
\usepackage{braket}
\journal{}
\begin{document}
\begin{frontmatter}
\title{Phase diagram for the spin-$3/2$ quantum  ferromagnetic Blume-Capel model in a  transverse crystal field: an approximation at the mean-field level}
\author{D. C. Carvalho$^1$, J. A. Plascak $^{2,3}$}
\address{$1$ Instituto Federal do Norte de Minas Gerais - Campus Salinas,
MG - CEP:39560-000 , Brazil
\\
$2$ Departamento de F\'\i sica, Instituto de Ci\^encias Exatas,
 Universidade Federal de Minas Gerais, C.P. 702, 
 30123-970 Belo Horizonte, MG - Brazil.\\
 $^3$ Center for Simulational Physics, University of Georgia, 30602 Athens,
Georgia, USA.
}
\begin{abstract}
We investigate the phase diagram for the spin-$3/2$ ferromagnetic Blume-Capel model in a transverse crystal field using the standard mean-field approximation within the framework of Bogoliubov inequality for free energy.
We draw a very rich phase diagram with first- and second-order transition lines; tricritical and tetracritical points; critical endpoint of order $2$ and double critical endpoint. Additionally, the behaviour of magnetisation as a function of temperature over a wide range of values of both longitudinal and transverse crystal fields is also analysed. To the best of our knowledge, this quantum spin model has only been studied employing an effective field theory, which in turn was not able to characterise completely the multicritical phenomena in its phase diagram, because that procedure is not based on free energy. Thus,
our findings on the phase diagram for the present model are novel as they have been not previously reported. 
\end{abstract} 
\end{frontmatter}
\section{Introduction}
\label{intro}
The investigation of lattice quantum spin models has attracted a lot of attention from researchers in  the realm of condensed-matter physics. 
Quantum spin models have not only been used to study magnetic properties of materials but mainly for the 
investigation of the effect of both thermal and quantum fluctuations in driving phase transitions \cite{sach,vojta}. 
One of the most emblematic example of quantum spin models is the spin-$1/2$ Ising model in an external transverse magnetic field, which has been successfully employed to examine quantum phase transition in the ionic crystal LiHoF$_4$ \cite{bitko}. Furthermore, this model has been solved exactly in one-dimension by Pfeuty \cite{pfeuty}.

However, for certain types of magnetic materials with effective spin quantum number $s>1/2$, especially rare-earth compounds \cite{ravi}, besides spin-spin  interactions in Ising-like models, there is also a significant interaction of spins with an internal crystal field \cite{abra}. 
Thus, to study such spin systems, one must add a crystal-field interaction in the Hamiltonian spin model. 
As a result, the spin-$1$ quantum Ising model with uniaxial or biaxial crystal field has been proposed in the 
literature \cite{strecka,review}. But, in general, spin models of spin quantum number  $s>1/2$ lacks integrability, and only approximate results may be feasible. In fact, the  spin-$1$ quantum Ising chain with uniaxial or biaxial crystal field is a rare example of an exactly solvable  quantum spin model with $s>1/2$ \cite{PRL,PRB-exact}. Additionally, quantum critical points have been exactly located in one-dimension, and  with high precision in two- and three-dimensional lattices, by means of a mapping onto spin-$1/2$ quantum Ising 
model in a transverse magnetic field \cite{Oitmaa}.
Also, for this model, phase diagram at finite temperatures has been studied by using analytical and numerical techniques: standard mean-field approximation \cite{J.Ric, Edde}, standard effective-field theory \cite{Zhang}, improved pair approximation \cite{diego}, effective-field theory with correlations \cite{costabile} and effective correlated mean-field scheme \cite{ze}. 

On the other hand, the phase diagram for corresponding
spin-$3/2$ quantum Ising model with biaxial  crystal
field or, equivalently,  spin-$3/2$ Blume-Capel model with a transverse crystal field (see Eq. (\ref{ham}))
has not yet been fully analysed in the literature. To the best of our knowledge, only  effective-field theory (EFT)
with self-spin correlations has been used to investigate the phase diagram of this model at finite temperatures on a honeycomb lattice \cite{jiang2003}. Additionally, even the phase diagram for spin-$3/2$ quantum Ising model with uniaxial crystal field (this corresponds to the limiting case $\Delta=0$ in Hamiltonian (\ref{ham}))
has only been explored by employing EFT on honeycomb, square and simple cubic lattices \cite{jiang2002}.
Nonetheless, EFT  is not suitable to identify first-order transitions lines since free energy is not available in this procedure, then the phase diagram reported on these models so far are not fully reliable. In fact, we have found novel interesting features in the phase diagram for the spin-$3/2$ Blume-Capel model with a transverse crystal field that has been neglected in the literature,  
 namely a presence of double critical endpoint; critical endpoint of order two; and tetracritical point. We have also located critical and tricritical points, besides first- and second-order transitions lines.
 
In this work, we go beyond the EFT at examining the details of the phase diagram for the spin-$3/2$ quantum Blume Capel model with a transverse crystal field by using the standard mean-field approximation (MFA) within the framework of Bogoliubov inequality.
The model under investigation is described by the following Hamiltonian
\begin{equation}
\label{ham}
{\cal H} = -J\sum_{<i,j>}S_{i}^{z}S_{j}^{z}-\Delta\sum_{i=1}^{N}{S_{i}^{z}}^{2}
-\Gamma\sum_{i=1}^N{S_{i}^{x}}^{2},
\end{equation}
where $J>0$ is the ferromagnetic exchange interaction between nearest-neighbour spins on the lattice sites $i$ and $j$;  the first sum counts all nearest-neighbour pairs $<i,j>$; $N$ is the total number of sites on the hypercubic lattice; $\Delta$ is the 
longitudinal crystal field and $\Gamma$ is the transverse crystal field; $S_i^\alpha$ 
$(\alpha=x,~y$ or $z$) are the spin operator components at site $i$, and  the set of eigenvalues of $S_i^z$ operator is $\{-3/2,-1/2,1/2,3/2\}$.

The present paper is organized as follows. In the next section, we describe the Bogoliubov variational approach and the Landau expansion for free energy.
In Section \ref{res}, we show the results and discusson on phase diagram for the limiting case ($\Delta=0$), and for the complete model in both
($\Delta/Jc,\frac{k_BT_c}{Jc}$)  and ($\Gamma/Jc,\frac{k_BT_c}{Jc}$) planes.
We close with some concluding remarks, and in the Appendix, we present some mathematical expressions  according to the analytical procedures employed herein.
\section{Variational approach for free energy}
\label{ham-var}
\subsection{Bogoliubov variational method}
We employ a variational method to study the present model. This approach is based on the Bogoliubov inequality        
\cite{falk}
\begin{equation} \label{bog}
F\leq F_{0}+\langle {\cal H}-{\cal H}_0(h)
\rangle_{0}\equiv \Phi(h) ,
\end{equation}
where ${\cal H}$ denotes the Hamiltonian (\ref{ham}) under study,
${\cal H}_0(h)$ represents a tentative Hamiltonian, which is exactly solvable,
and it is a function of variational parameters labeled by $h$. $F$ is the free 
energy of the model described by $\cal H$, $F_0$ is the free energy related to the
tentative Hamiltonian ${\cal H}_0$, and $<... >_0$ denotes the average that is taken over the canonical ensemble related to the system defined  by ${\cal H}_0$ in thermal equilibrium with a heat reservoir.

To begin we describe this variational scheme briefly below.
First of all, we have chosen a tentative Hamiltonian at the mean-field level: 
\begin{equation}
 \label{free-ham-bc-3/2}
{\cal H}_{0}=-{\sum_{i=1}^{N}}\left[hS_{i}^{z}+\Gamma{S_{i}^{x}}^{2}+\Delta{S_{i}^{z}}^{2}\right],
\end{equation}
where $h$ denotes the variational parameter, and the sum is over all spins. By rewriting that expression into a matrix form (if the reader is interested in this analytical part, check the corresponding matrix and its eigenvalues in  Appendix), the resulting $4\times4$ matrix is diagonalised in a straightforward way yielding the following closed-form for the partition function ${\cal Z} _ {0} $:
\begin{equation}
 {\cal Z} _ {0}=2e^{\frac{5\beta(\Delta+\Gamma)}{4}}\left[e^{\frac{\beta{h}}{2}}\cosh\left(\frac{\beta{x}}{2}\right)+
e^{\frac{-\beta{h}}{2}}\cosh\left(\frac{\beta{w}}{2}\right)\right],
\end{equation}
where 
\begin{equation}
 x = \sqrt{[-2(h+\Delta)+\Gamma]^2+3\Gamma^{2}},
\end{equation}
\begin{equation}
 w = \sqrt{[-2(h-\Delta)-\Gamma]^2+3\Gamma^{2}}.
\end{equation}
From the partition function, ${\cal Z} _ {0} $, we can find out the analytical expression for  
the approximate magnetisation per site, $m$, which depends upon variational parameter $h$, as follows: 
\begin{eqnarray}
 \label{mag-3/2}
m=\frac{1}{\beta}\frac{\partial{\ln{\cal Z}_{0}}}{\partial{h}}=\bigg\{
e^{\frac{\beta{h}}{2}}\left[\cosh\left(\frac{\beta{x}}{2}\right)-2\sinh\left(\frac{\beta{x}}{2}\right)\left(\frac{-2(h+\Delta)+\Gamma}{x}\right)\right]
\nonumber\\
-e^{-\frac{\beta{h}}{2}}\left[\cosh\left(\frac{\beta{w}}{2}\right)+2\sinh\left(\frac{\beta{w}}{2}\right)\left(\frac{-2(h-\Delta)-\Gamma}{w}\right)\right]
\bigg\}{\bigg/}\nonumber\\
2\left[e^{\frac{\beta{h}}{2}}\cosh\left(\frac{\beta{x}}{2}\right)+
e^{\frac{-\beta{h}}{2}}\cosh\left(\frac{\beta{w}}{2}\right)\right].
\end{eqnarray}
We can now obtain the function $\Phi(h)$:
\begin{equation}
\label{fi-32}
 \Phi(h)=-\frac{JNcm^2}{2}+Nhm-Nk_{B}T\ln{\cal Z}_{0},
\end{equation}
where $c$ is the coordination number of the hypercubic lattice. By minimising the function $\Phi$ with respect to the  variational parameter, $h$, we arrive at the following relation
\begin{equation}
\label{par-3/2}
 \frac{h}{Jc}=m,
\end{equation}
which must be used to obtain an analytical approximate expression for free energy per particle:
\begin{eqnarray}
\label{free-energy-3/2}
\frac{\Phi_{min}}{N}=\frac{Jcm^2}{2}-k_{B}T\ln{\cal Z}_{0}.
\end{eqnarray}
The thermodynamic properties of the model can be achieved by solving numerically Eqs. (\ref{mag-3/2}) and (\ref{par-3/2}). For known input values of the dimensionless fields, $\frac{\Gamma}{Jc}$ and $\frac{\Delta}{Jc} $, the reduced variational parameter, $\frac{h}{Jc}$, can be obtained self-consistently as a function of reduced temperature, $\frac{k_ {B} T} {Jc}$, and then the thermodynamic quantities can be computed.

In addition, if we let the variational parameter goes to zero ($h\rightarrow0$), magnetisation also 
goes to zero continuously, which characterizes a continuous transition from  ferromagnetic ($m\neq0$) to  paramagnetic phase ($m=0$). Hence, one may analyse directly the critical region of the model by equating Eqs. (\ref{mag-3/2}) and (\ref{par-3/2}) in the limit $h\rightarrow0$. As a result, the reduced critical temperature, $\frac{k_B{T_c}}{Jc}$,  can be calculated from the following equation: 
\begin{eqnarray}
\label{cri-32}
 \frac{k_B{T_c}}{Jc}&=&\frac{1}{4}\left\{1-\frac{4b}{x(0)}\tanh\left(\frac{\beta_{c}{x(0)}}{2}\right)+\frac{4b^2}{x^2(0)}
+\frac{8\tanh\left(\frac{\beta_{c}{x(0)}}{2}\right)}{x(0)\beta_c}\left[1-\frac{b^2}{x^2(0)}\right]\right\},\nonumber\\
\end{eqnarray}
where 
$$\beta_c\equiv{\frac{1}{k_BT_c}},$$
$$ b = -2\Delta+\Gamma,$$
$$ x(0)\equiv{x(h\rightarrow0)}=\sqrt{b^2+3\Gamma^2}.$$
In general, the above critical equation can only be evaluated numerically, for example, by using the Newton-Raphson technique. On the other hand, since the present method is based on free energy, first-order transition lines may also be accurately characterised by comparing free energies of different phases.

The critical equation (\ref{cri-32}) takes simpler analytical form in certain limiting cases:
\\
({\it i})  For $\Gamma=0$ and $\Delta\neq0$, the classical spin-$3/2$ Blume-Capel model is obtained, and then the critical lines are determined by the following critical equation:
\begin{equation}
 \frac{k_BT_c}{Jc}=\frac{5}{4}+\tanh\left(\beta_c\Delta\right).
\end{equation} 
This classical model has been studied extensively in the literature (see, e.g., Refs. \cite{plascak,costabile2013} and references therein).
\\
({\it ii}) For $\Delta=0$ e $\Gamma\neq0$, the spin-$3/2$ Ising model in a transverse crystal field - which is a quantum model - is obtained, and critical temperature can be calculated as a function of $\Gamma/J$, as follows:
\begin{equation}
 \frac{k_BT_c}{Jc}=\frac{1}{4}\left[2-\frac{\tanh(\beta_c\Gamma)}{\Gamma/J}\left(\frac{2\Gamma}{J}-\frac{3k_BT_c}{J}\right)\right].
\end{equation}
\subsection{Landau Free Energy}
\label{TCP-bc-32}
To investigate the presence of tricritical points in the phase diagram, we have made a typical Landau expansion for the function $\Phi$, which is given by Eq.(\ref{fi-32}). According to this scheme,  the function $\Phi$ has been expanded as a power series in the order parameter $m$:
\begin{eqnarray}
\label{le-32}
\frac{\Phi}{NJc}&=&\Phi_{0}+\frac{1}{2}a_{2}\left(\frac{k_BT}{Jc},\frac{\Gamma}{Jc},\frac{\Delta}{Jc}\right)m^{2}+
\frac{1}{4}a_4\left(\frac{k_BT}{Jc},\frac{\Gamma}{Jc},\frac{\Delta}{Jc}\right)m^4\nonumber\\
&+& O(m^6),
\end{eqnarray}
where both coefficients $a_2$ and $a_4$ depends upon 
longitudinal and transverse crystal fields, besides temperature. Tricritical points are located whenever the following conditions are satisfied:
\begin{eqnarray}      
 a_{2}\left(\frac{k_BT}{Jc},\frac{\Gamma}{Jc},\frac{\Delta}{Jc}\right)=0~~~~~~~~\mbox{and}~~~~~~~
a_{4}\left(\frac{k_BT}{Jc},\frac{\Gamma}{Jc},\frac{\Delta}{Jc}\right)=0.~~~~~
\nonumber 
\end{eqnarray}          
Analytical expressions for these coefficients are available in the Appendix for the interested reader. 
\section{Results and Discussion}
\label{res}
In this section, we show numerical results for the spin-$3/2$ Blume-Capel model in a transverse crystal field. The behaviour of magnetisation and the details of phase diagrams are analysed as a function of the dimensionless fields, $\Gamma/Jc$ and $\Delta/Jc$, and the reduced temperature, $\frac{k_BT}{Jc}$. As we shall see below, a remarkably rich phase diagram has been drawn. 
\subsection{Magnetisation}
\begin{figure}[h!]
\centering
\includegraphics[width=1.\textwidth]{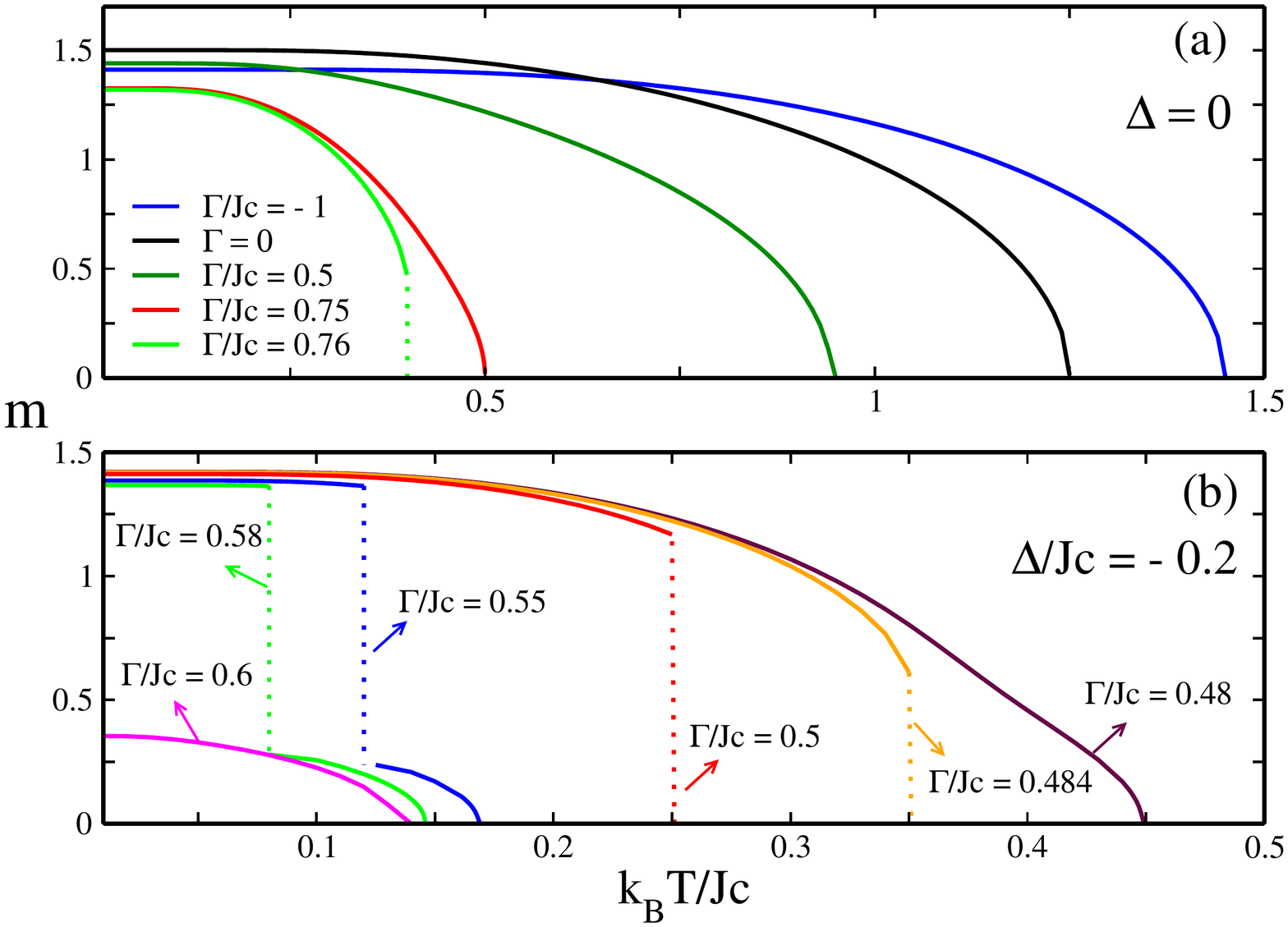}
\caption{(color online). Magnetisation, $m$, as a function of reduced temperature, $t={k_{B}T}/{Jc}$, for 
 (a) $\Delta=0$ and (b) $\Delta/Jc=-0.2$, and several values of transverse crystal field
 $\Gamma/Jc$.}\label{fig-mag-bc-32}
\end{figure}

\enlargethispage{1cm}
As shown in Figure \ref{fig-mag-bc-32}, the behaviour of magnetisation, $m$, modifies substantially as the dimensionless fields, $\Delta/Jc$ and $\Gamma/Jc$, are varied. In fact, the character of phase transition changes from second- to first-order transition (dotted lines). Fig. \ref{fig-mag-bc-32} (a) depicts a special limiting case: spin-$3/2$ Ising model in a transverse crystal field ($\Delta=0$). In this case, one notes that for $\Gamma>0$, the magnetisation does not go to zero continuously for each value of that field. For example, for
$\Gamma/Jc=0.76$, the magnetisation jumps to zero, which characterises a first-order transition from the ferromagnetic to paramagnetic phase. In addition, for $0\leq\Gamma/Jc\leq0.75$,  the critical temperature  decreases with increasing transverse crystal field due to enhanced quantum fluctuations, which in turn makes difficult the spin ordering along the $z$-axis. 

By contrast, as illustrated for 
$\Gamma/Jc=-1$, only continuous transition lines have been found for $\Gamma<0$. We would like to draw the attention of the reader to the critical temperature for $\Gamma/Jc=-1$ that is notably lower than $\Gamma=0$ (classical Blume-Capel), although quantum fluctuations are absent in classical limit. 
This apparent paradox can be resolved if one observes that only two quadratic spin operator components are independent,
since $S^2={S_i^{x}}^{2}+{S_i^{y}}^{2}+{S_i^{z}}^{2}$, and $S^2$ is a constant. In fact,  
we can rewrite the Hamiltonian (\ref{ham}) as 
a function of $S_{i}^{z}, {S_{i}^{z}}^{2}$
and ${S_{i}^{y}}^2$:
\begin{equation}
\label{ham-eff}
{\cal H} = -J\sum_{<i,j>}S_{i}^{z}S_{j}^{z}-\Delta_{eff}\sum_{i=1}^{N}{S_{i}^{z}}^{2}
+\Gamma\sum_{i=1}^N{S_{i}^{y}}^{2}-\Gamma{N}S^2,
\end{equation}
where we have defined an effective longitudinal field,
$\Delta_{eff}=\Delta-\Gamma$, which depends upon both
crystal fields $\Delta$ and $\Gamma$. 
Therefore, negative transverse crystal field increases the effective longitudinal  field with respect to the case 
$\Gamma = 0$, even though quantum fluctuations have indeed been enhanced. This explains why the critical temperature for $\Gamma/Jc=-1$ is greater than for $\Gamma/Jc=0$. 

To be more precise, the model described by Eq. (\ref{ham}) is able to host four types of ferromagnetic phases:  two of them have magnetisations close to $m=\pm3/2$ (labeled herein by $O_{3/2}$), and the other two have magnetisations close to $m=\pm1/2$  (labeled herein by $O_{1/2}$), in which these values of magnetisation are considered at absolute zero\footnote {For $\Gamma=0$, at absolute zero, the magnetisation for $O_{3/2}$
and $O_{1/2}$ phases is exactly equal to $m=\pm3/2$ and $m=\pm1/2$, respectively, since quantum fluctuations are absent in this case. Phases with negative magnetisations ($m=-1/2$ and $m=-3/2$) are also present due to symmetry of the Hamiltonian (\ref{ham}) regarding the inversion of all spins along the $z$-axis.}. In addition
to ferromagnetic-paramagnetic phase transitions, there are also, in low temperatures region, transitions between ordered phases as illustrated in Fig. \ref{fig-mag-bc-32} (b) for $\Delta/Jc=-0.2$. In fact, this spin model undergoes a first-order transition between
$O_{3/2}$ and $O_{1/2}$ phases for $0.55\leq\Gamma/Jc\leq0.58$. 
By contrast, upon increasing transverse crystal field, the $O_{3/2}$ phase becomes unstable (see 
the curve for  $\Gamma/Jc=0.6$), and the magnetisation goes continuously to zero, which characterises a second-order transition from $O_{1/2}$ to paramagnetic phase.  
However, for $0.484\leq\Gamma/Jc<0.55$ one notes a first-order transition between the $O_{3/2}$ phase and paramagnetic one, whereas continuous transitions between these phases occurs for $\Gamma/Jc<0.48$. 
\subsection{Limiting case: Spin-$3/2$ Ising Model in a Transverse Crystal Field}
%
\begin{figure}[ht]
\begin{center}
\includegraphics[width=1.\textwidth]{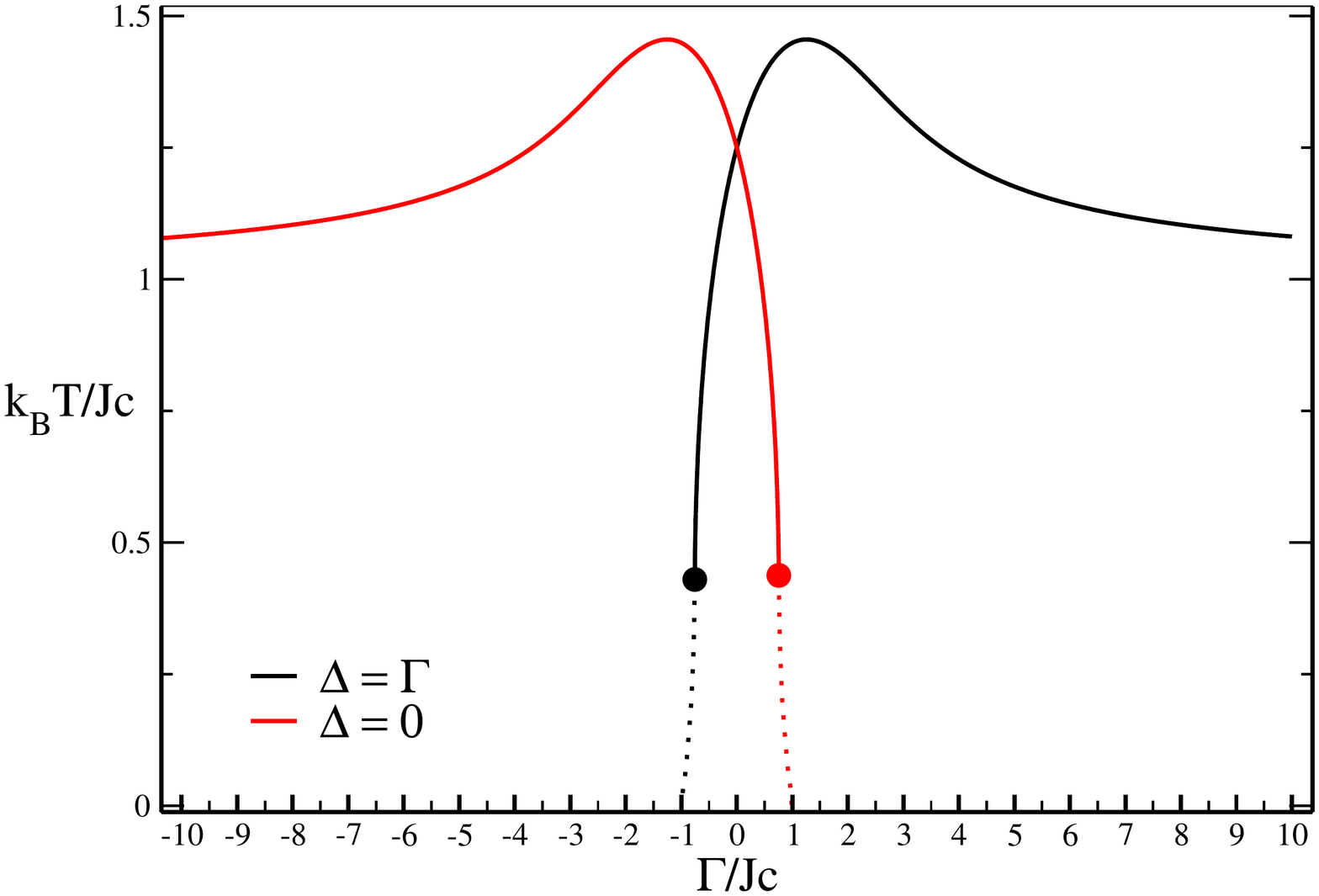} 
\caption{\label{pha-Isi-32}
(color online). 
Phase diagram for spin-$3/2$ Ising model in a transverse crystal field. 
The continuous lines refer
to second-order phase transitions, and  the dotted lines refer to first-order phase transitions. The circles represent tricritical points. }
\end{center}
\end{figure}
%
%

The phase diagram for spin-$3/2$ Ising model in a transverse crystal field is shown in Fig. \ref{pha-Isi-32}. As expected for this model, no quantum critical point has been found by  using the present mean-field approach. According to
Oitmaa and Brasch \cite{Oitmaa}, the ground state of this model can be exactly mapped onto the spin-$1/2$ Ising model with both longitudinal and transverse magnetic field, which in turn does not exhibit quantum phase transition. Therefore MFA predicts correctly such a result. 

In addition to that, a tricritical point has been found for $\Delta=0$ and $\Delta=\Gamma$ at finite temperatures. The former case corresponds to the application of transverse field along the $x$-axis, while the latter one is equivalent to the application of transverse field along the $y$-axis, as shown by Eq.(\ref{ham-eff}). Regardless of the direction of the transverse crystal field, the phase diagrams are similar: the transition curves are symmetric with respect to the operation $\Gamma\rightarrow-\Gamma$. Moreover, 
for $\Gamma\rightarrow-\infty$ and  $\Delta=0$, or similarly, for  $\Gamma\rightarrow\infty$ and $\Delta=\Gamma$, the critical temperature approaches an asymptotic value of                
  $\frac{k_BT_c}{Jc}=1$. Consequently, in thsese limits, the spins order down to absolute zero, even if the transverse crystal field is very strong. One also notes that both curves exhibt a maximum at the points: $(-1.25,1.455)$ for $\Delta=0$, and 
$(1.25,1.455)$ for $\Delta=\Gamma$. As discussed in  previous section, the reason of this maximum in phase diagram can be
understood as a consequence of the fact that 
only two quadratic spin operator components are independent, besides the non-commutativity of spin operators. 

Also, first-order transition line from the ferromagnetic to paramagnetic phase has also been located. 
It should be pointed out that previous work by using EFT was not able to identify  first-order phase transition lines \cite{jiang2002}. However, in that reference, tricritical points were reported on both square and simple cubic lattices, and a comparison of those data with our findings is depicted in Table \ref{tab-32}.
\begin{table}
\caption{\it Tricritical points $\left(\frac{\Gamma_T}{J},\frac{k_BT_T}{J}\right)$ 
of spin-$3/2$ Ising model in a transverse crystal field ($\Delta=0$) on the square ($c=4$) and simple cubic ($c=6$) lattices, according to the present mean-field approach (MFA) and effective-field theory (EFT)\cite{jiang2002}.
}
\label{tab-32}
\vskip0.1in
\begin{center}
\begin{tabular}{ccc} 
\hline \hline 	
       &MFA    & EFT\\ 
\hline \hline 
$c=4$ &$(3.028,1.751)$  & $(3.159,1.037)$ \\
$c=6$ &$(4.542,2.626)$ & $ (4.630,1.918) $\\
\hline \hline 
\end{tabular}
\end{center}
\end{table}
%
\subsection{Phase Diagram in the ($\Delta/Jc$, $\frac{k_BT_c}{Jc}$) plane.}
\begin{figure}[ht]
\begin{center}
\includegraphics[width=1.\textwidth]{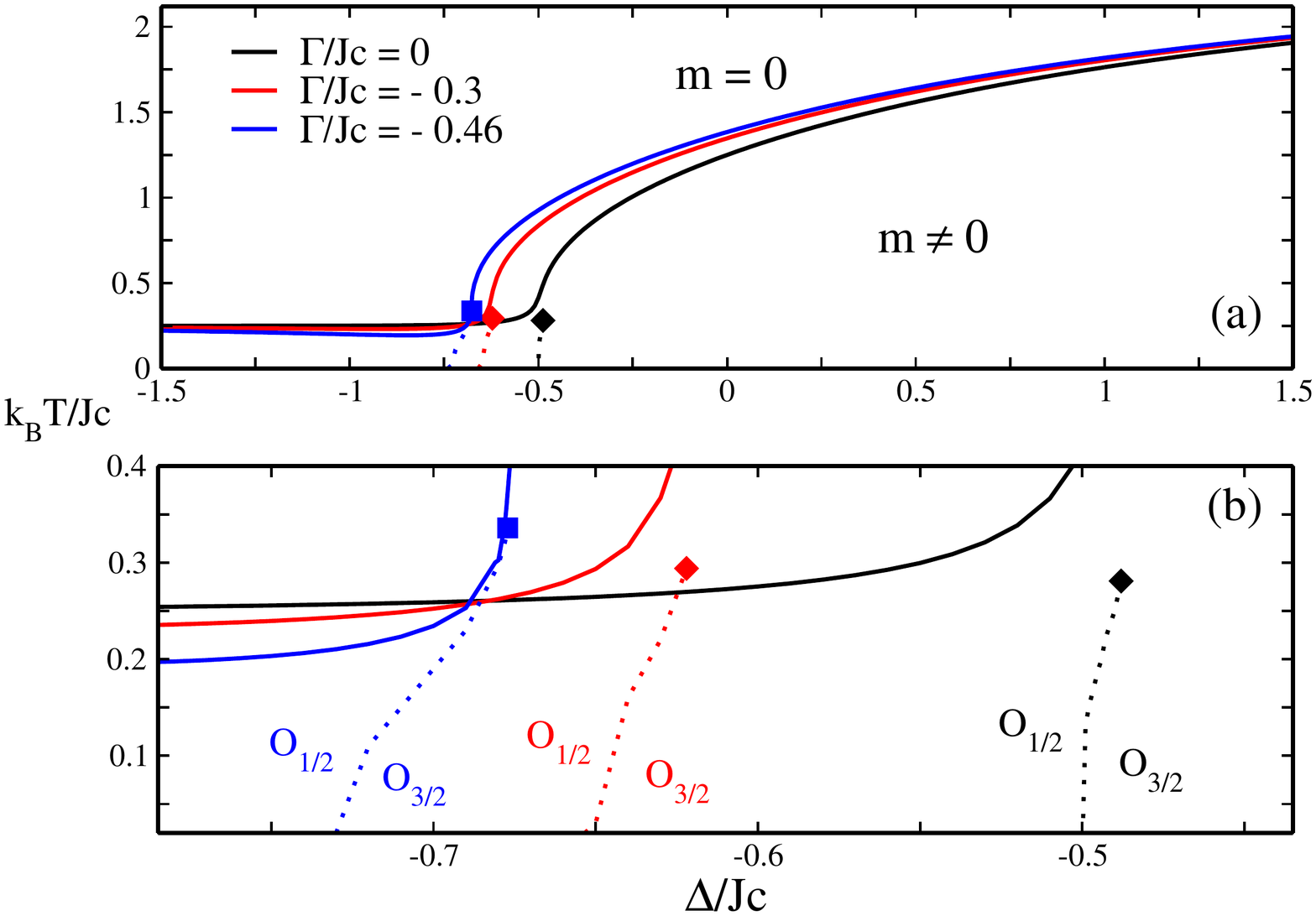} 
\caption{\label{phadel}
(color online). (a) Phase diagram in the 
($\Delta/Jc,\frac{k_BT_c}{Jc}$) plane for some values of $\Gamma/Jc$. The continuous lines refer to second-order phase transitions and the dotted lines refer to first-order transitions between $ O_ {3/2} $ and $ O_ {1/2} $ ordered phases.
The diamonds represent double critical endpoints, and the square denotes a tetracritical point. (b)
Low temperatures region on a finer scale.}
\end{center}
\end{figure}

It is a commonplace that classical spin-$3/2$ Blume-Capel model does not exhibit a tricritical point \cite{alcaraz,pla-32, plascak,lara1,costabile2013}. Moreover, it is well known that in low temperatures there is a line of coexistence of four ordered phases, which ends up at a double critical endpoint. At this special point, two critical phases coexist: $ O_ {3/2}~(m=3/2)\equiv O_{1/2}~(m=1/2)$ and  $ O_ {3/2}~(m=-3/2)\equiv O_{1/2}~(m=-1/2)$. For completeness, this classical limiting case is showed in Figure \ref{phadel} for $\Gamma=0$. 

On the other hand, if one allows a finite transverse crystal field, quantum fluctuations are introduced by the non-commutativity of quantum spin operators and give rise to a much rich phase diagram. In fact, the double critical endpoint approaches the line of continuous transition for  
$-0.46<\Gamma/J<0$, and specifically at $\Gamma/Jc=-0.46$ it touches the curve and results in a tetracritical point. At this specific point,  four ordered phases become critical:
 $O_{3/2}~(m=\pm3/2)\equiv O_{1/2}~(m=\pm1/2$).
%
\begin{figure}[ht]
\begin{center}
\includegraphics[width=1.\textwidth]{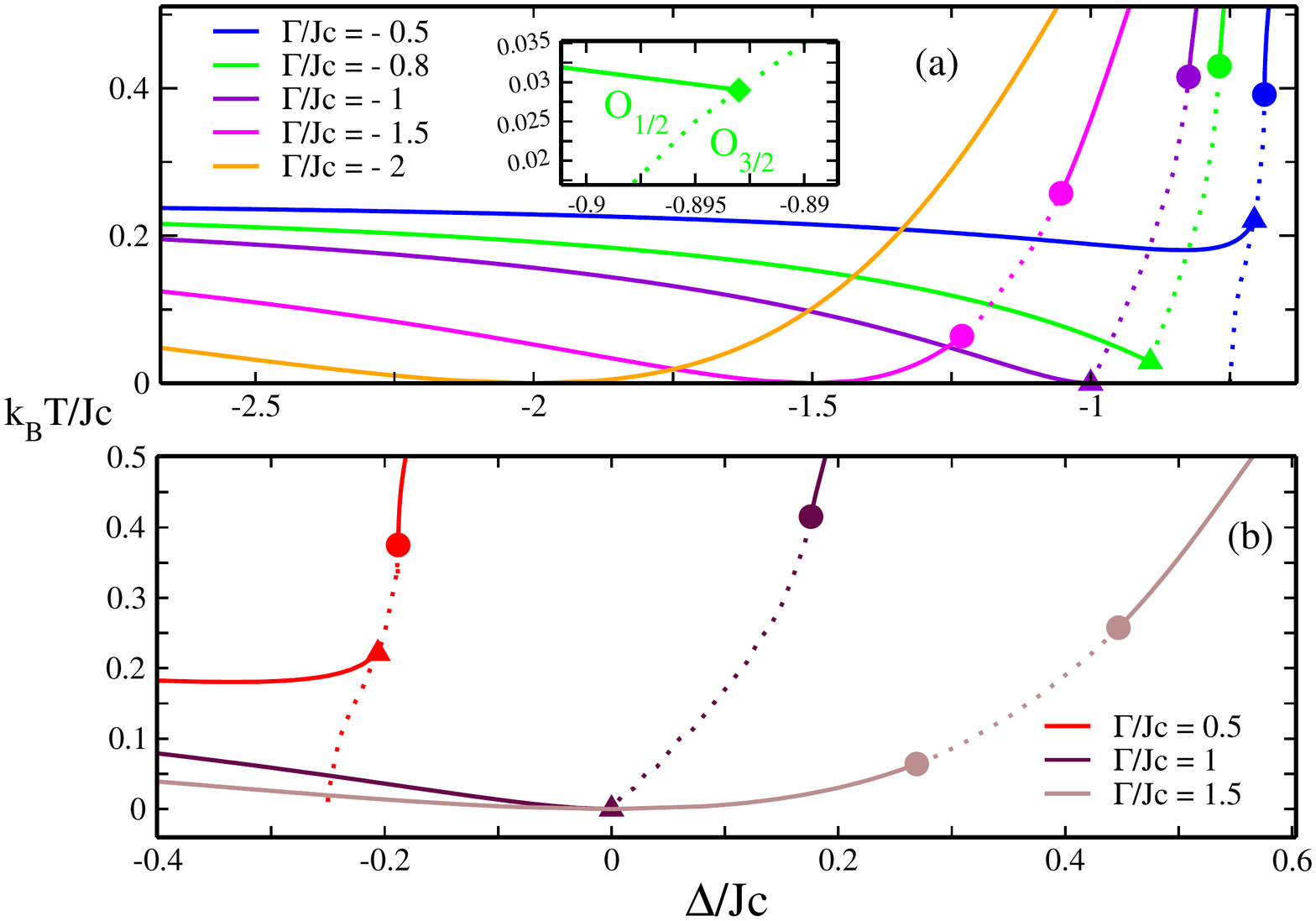} 
\caption{\label{pha-cont}
(color online). (a) Phase diagram in the ($\Delta/Jc,\frac{k_BT_c}{Jc}$) plane
for some negative values of $\Gamma/Jc$. The continuous lines refer to second-order phase transitions and the dotted lines refer to first-order transitions
The circles and triangles denote tricritical and critical endpoint of order $2$, respectively. The insert shows the curve for $\Gamma/Jc=-0.8$ in low temperatures region on a finer scale. (b) The same as (a) for $\Gamma/Jc>0$.}
\end{center}
\end{figure}
%
\\
In Figure \ref{pha-cont} (a), one notes that by increasing the magnitude of the transverse crystal field, for $-0.5\leq\Gamma/Jc\leq-1$, a tricritical point emerges in the phase diagram while the line of continuous transition between $O_{1/2}$ and paramagnetic phases ends up at a critical endpoint of order $2$.
\footnote{We use this nomenclature to distinguish critical endpoints according to the number of ordinary phases coexisting at that point. For example, at a critical endpoint of order $ n $, two ordinary phases become critical in the presence of $ n $ ordinary phases.}.
Therefore, the two $ O_ {1/2} $ phases  become critical in the presence of the two $ O_ {3/2} $ ordinary phases. Additionally, this critical endpoint moves to the low temperatures region, and finally reaches absolute zero at 
$\Gamma / Jc = -1 $.

For $ -1.1 \leq \Gamma / Jc \leq-1.5 $, two tricritical points appear in the phase diagram, as illustrated for $\Gamma / Jc = -1.5 $.
However, for $ \Gamma / Jc \leq-1.6 $, only continuous phase transitions have been found, as exemplified for $ \Gamma / Jc = -2 $.
Similar phase diagram is presented in Figure \ref{pha-cont} (b) for $\Gamma/Jc>0$.  
\subsection{Phase Diagram in the $(\Gamma/Jc $,$\frac{k_BT}{Jc})$ plane.}

\begin{figure}[h!]
\begin{center}
\includegraphics[width=1.\textwidth]{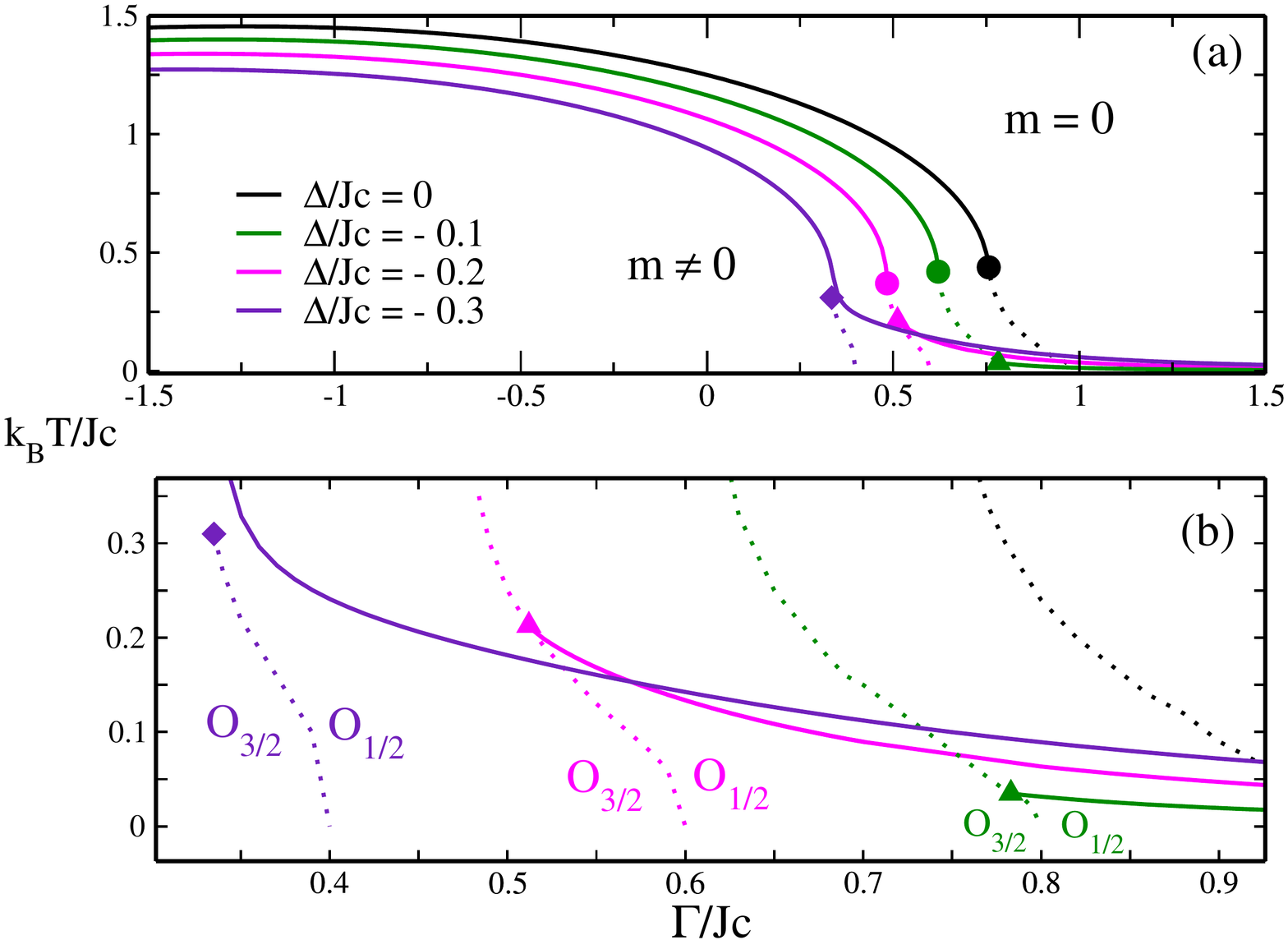} 
\caption{\label{phaxgam}
(color online). (a) Phase diagram in the $(\Gamma/Jc,\frac{k_BT}{Jc})$ plane
for some values of $\Delta/Jc$. The continuous lines refer to second-order phase transitions and the dotted lines refer to first-order transitions.
The triangules and the diamond denote critical endpoints of order $2$, and a double critical endpoint, respectively.
 (b) Low temperatures region on a finer scale.}
\end{center}
\end{figure}

In this section, we turn our attention to the phase diagram in the $(\Gamma/Jc,\frac{k_BT}{Jc})$ plane for various values of $\Delta/Jc$. 
As illustrated in Figure \ref{phaxgam}, the resulting phase diagram  exhibits a tricritical point besides a coexistence line of $ O_ {3/2} $ and paramagnetic phases for $ -0.3 <\Delta / Jc \leq0 $. One can also see that
the critical endpoint, which is absent for $\Delta=0$, appears in the phase diagram as the end-point of the continuous transition line between
$ O_ {1/2} $ and paramagnetic phases.
As $ \Delta / Jc $ decreases,
the critical endpoint moves towards the tricritical ones, and becomes a double critical endpoint at $ \Delta / Jc = -0.3 $. Similar behaviour is found over a wide range of
$\Delta/J$.

It should be mentioned that only first-order phase transitions have been found at absolute zero temperature for finite transverse crystal field. Furthermore, for $\Delta=0$ these transitions occur between the magnetically ordered phase ($m\neq0$) and paramagnetic one ($m=0$). By contrast, for  
$-0.3\leq\Delta/Jc\leq-0.1$ the spin model undergoes a first-order transition between $O_{3/2}$ and $O_{1/2}$ ordered phases.  As $\Gamma\rightarrow\infty$, the lines of critical points asymptotically approach absolute zero.
%
\section{Concluding Remarks}
\label{con}
We have investigated the  spin-$3/2$ Blume Capel model with a transverse crystal field at the mean-field level. We have seen that the presence of quantum fluctuations besides thermal ones give rise to a very rich phase diagram. We have characterised first-order transitions between $ O_ {3/2} $ and $ O_ {1/2} $ ordered phases as well as between ordered and paramagnetic ones. We have also identified tricritical points using a Landau expansion; double critical endpoints; critical endpoints of order two, and tetracritical points. It should be emphasised 
that previous study on the present model by employing EFT \cite{jiang2002,jiang2003}  was not able to capture the richness of details of the phase diagram that we have drawn, mainly because
free energy is not available in that scheme. 

Even though the present approach is still a mean-field approximation, we believe that our results 
may be accurate at least on the simple cublic lattice since quantum and thermal fluctuations are not so intense as would be, for example, in one-dimension or on spin-half models. Moreover, within the framework of Bogoliubov inequality, we correctly predict the absence of quantum critical points in the limiting case $\Delta=0$, which corresponds to the spin-$3/2$ Ising model with a transverse crystal field.  

As a final remark, the application of numerical procedures to the present model would be very welcome to be compared with our
analytical findings. Quantum Monte Carlo simulations (see, e.g., Ref.\cite{georgescu2014} and references therein), for example, could be employed as long as the quantum spin system under investigation is not frustrated.

\appendix
\section{}

By writing the ${\cal H}_{0}$ as a matrix (size $4\times4$) using the basis of eigenstates of operator $S^{z}$: 
 $\{|+3/2\rangle,|+1/2\rangle,$ $|-1/2\rangle,|-3/2\rangle\}$, we obtain
\[ {\cal H}_{0} = \left[\begin{array}{cccc}
-\left(\frac{3h}{2}+\frac{9\Delta}{4}+\frac{3\Gamma}{4}\right) & 0 & -\frac{\sqrt{3}\Gamma}{2}&0 \\
0 & -\left(\frac{h}{2}+\frac{\Delta}{4}+\frac{7\Gamma}{4}\right) & 0 &-\frac{\sqrt{3}\Gamma}{2} \\
-\frac{\sqrt{3}\Gamma}{2} & 0 & \left(\frac{h}{2}-\frac{\Delta}{4}-\frac{7\Gamma}{4}\right)&0\\
0  & -\frac{\sqrt{3}\Gamma}{2} & 0 & \left(\frac{3h}{2}-\frac{9\Delta}{4}-\frac{3\Gamma}{4}\right)
\end{array} \right], \] 
whose eigenvalues are
\begin{eqnarray}
\lambda_{1,2}&=&\frac{1}{2}\left(-h_1-\frac{5\Delta}{2}-\frac{5\Gamma}{2}\pm\sqrt{[-2(h+\Delta)+\Gamma]^2+3\Gamma^{2}}\right),\nonumber\\
\lambda_{3,4}&=&\frac{1}{2}\left(h_1-\frac{5\Delta}{2}-\frac{5\Gamma}{2}\pm\sqrt{[-2(h-\Delta)-\Gamma]^2+3\Gamma^{2}}\right).
\end{eqnarray}

We also show below analytical expressions for  coefficients $a_2$ and $a_4$ in the Landau expansion given by Eq. (\ref{le-32}): 
\begin{eqnarray}
 a_2&=&-\frac{1}{2}\left[1+\frac{1}{C-F\tanh\left(\frac{\beta{J}c{x(0)}}{2}\right)}\right],
\end{eqnarray}
\begin{eqnarray}
a_4&=&\frac{5}{4}\left[\frac{\frac{\beta{J}c}{2}\left(C-F\tanh\left(\frac{\beta{J}cx(0)}{2}\right)\right)^{2}
+\frac{1}{6}\left(A-B\tanh\left(\frac{\beta{J}c{x(0)}}{2}\right)\right)}{\left(C-F\tanh\left(\frac{\beta{J}c{x(0)}}{2}\right)\right)^{4}}\right],\nonumber\\
\end{eqnarray}
where
\begin{eqnarray}
C&=&-\beta{Jc}\left(\frac{1}{4}+\frac{b^2}{{x(0)}^2}\right); \nonumber
\end{eqnarray}
\begin{eqnarray}
 F&=&-\beta{J}c\frac{b}{x(0)}+\frac{2}{x(0)}\left(1-\frac{b^2}{{x(0)}^2}\right);\nonumber
\end{eqnarray}
\begin{eqnarray}
 A&=&(-\beta{Jc})^{3}\left(\frac{1}{16}+\frac{3}{2}\frac{b^2}{{x(0)}^2}+\frac{b^4}{{x(0)}^4}\right)
+12\frac{(\beta{Jc})^2}{x(0)}\left(\frac{b}{x(0)}-\frac{b^3}{{x(0)}^3}\right)\nonumber\\
&+&48\beta{Jc}\frac{b}{{x(0)}^3}\left(\frac{b}{x(0)}-\frac{b^3}{{x(0)}^3}\right)
-12\frac{\beta{J}c}{{x(0)}^2}\left(1-\frac{b^2}{{x(0)^2}}\right)^2;\nonumber
\end{eqnarray}
\begin{eqnarray}
B&=&(-\beta{Jc})^3\left(\frac{b}{2x(0)}+\frac{2b^3}{{x(0)}^3}\right)+12\frac{(\beta{Jc})^2}{x(0)}\left(\frac{1}{4}+\frac{3b^2}{4{x(0)}^2}
-\frac{b^4}{{x(0)}^4}\right)\nonumber\\
&+&24\beta{Jc}\frac{b}{{x(0)}^3}\left(1-\frac{b^2}{{x(0)}^2}\right)-\frac{24}{{x(0)}^3}\left(1+\frac{5b^4}{{x(0)}^4}-\frac{6b^2}{{x(0)}^{2}}\right);
\nonumber
\end{eqnarray}
with
$$ b = -2\frac{\Delta}{Jc}+\frac{\Gamma}{Jc},$$

$$ x(0)=\sqrt{b^2+3\left(\frac{\Gamma}{Jc}\right)^2}.$$

\section{Acknowledgments}
The authors thank CNPq and 
CAPES (Brazilian agencies) for the financial support.
\section{References}

\end{document}